\documentclass[%
 reprint,
superscriptaddress,
 amsmath,amssymb,
 aps,
]{revtex4-2}

\usepackage{graphicx}
\usepackage{dcolumn}
\usepackage{bm}
\usepackage{xcolor}

\begin{document}

\preprint{APS/123-QED}

\title{Dynamics of phase defects trapped in optically imprinted orbits \\ in dissipative binary polariton condensates}

\author{Jan Wingenbach}
 \affiliation{Department of Physics and Center for Optoelectronics and Photonics Paderborn (CeOPP), Universit\"{a}t Paderborn, Warburger Strasse 100, 33098 Paderborn, Germany}
 
\author{Matthias Pukrop}%
\affiliation{Department of Physics and Center for Optoelectronics and Photonics Paderborn (CeOPP), Universit\"{a}t Paderborn, Warburger Strasse 100, 33098 Paderborn, Germany}%

\author{Stefan Schumacher}
\affiliation{Department of Physics and Center for Optoelectronics and Photonics Paderborn (CeOPP), Universit\"{a}t Paderborn, Warburger Strasse 100, 33098 Paderborn, Germany}%
\affiliation{Wyant College of Optical Sciences, University of Arizona, Tucson, AZ 85721, USA}%

\author{Xuekai Ma}
\affiliation{Department of Physics and Center for Optoelectronics and Photonics Paderborn (CeOPP), Universit\"{a}t Paderborn, Warburger Strasse 100, 33098 Paderborn, Germany}%


\begin{abstract}
We study the dynamics of phase defects trapped in a finite optically imprinted ring lattice in binary polariton condensates, under the influence of the cross-interaction (CI) between the condensates in different spin components and the spin-orbit interaction (SOI). In this configuration, we find that a vortex circulates unidirectionally in optically induced orbits because of the Magnus force acting in the polariton fluid, and the vortex' angular velocity is influenced by the SOI and CI. Interestingly, in our system, these two interactions can also lead to elongated and frozen phase defects, forming a frozen dark solution with similarity to a dark soliton but with finite size in both spin components. When the dark solution is stretched further to occupy the entire orbit of a condensate ring, the phase defect triggers a snake instability and induces the decay of the dark ring solution. The circulation direction of a single vortex is determined by the Magnus force. This situation is more complex for the group motion of multiple vortices because of significant vortex-antivortex interaction. The collective motion of such vortex constellations, however, can be determined by the SOI.
\end{abstract}

\maketitle


\section{Introduction}
Vortices carrying quantized orbital angular momenta have attracted significant attention in a variety of physical systems such as in nonlinear optics~\citep{kivshar2003optical},  superconductors~\citep{blatter1994vortices}, atomic condensates~\citep{fetter2009rotating}, and polariton condensates~\citep{lagoudakis2008quantized,lagoudakis2009observation,roumpos2011single,ma2020realization}. Polaritons are quasiparticles, formed due to the strong coupling of photons and excitons in planar semiconductor microcavities. As half-light half-matter hybrid particles, they can show macroscopic coherence or condensation~\citep{deng2002condensation,kasprzak2006bose} under nonresonant optical excitation and possess strong nonlinearity. Different from the formation mechanism of vortices in atomic condensates~\citep{fetter2009rotating}, the driven-dissipative nature and short lifetime of polariton condensates enable the survival of phase defects during the pumping, evolving spontaneously into vortices. These spontaneously formed vortices can be trapped inside local potential valleys~\citep{lagoudakis2009observation}. The nonresonant optical excitation can also induce a repulsive potential energy landscape resulting from the interactions with the induced exciation reservoir. A periodically modulated pump profile imprints a periodic potential and confines the motion of the vortices. In this case, even unstable one-dimensional phase defects, known as dark solitons, can be stabilized~\citep{PhysRevLett.118.157401}. However, dark soliton stripes in two dimensions are unstable and they finally evolve into vortex-antivortex pairs because of the snake instability, which has been widely reported in both atomic (conservative)~\citep{PhysRevLett.86.2926,PhysRevLett.90.120403} and polariton (dissipative) condensates~\citep{PhysRevB.89.235310,PhysRevLett.123.215301,Claude:20}. The decay of the two-dimensional dark solitons can be suppressed by supersonic flow of particles, which is known as oblique dark solitons formed  in the wake of an obstacle that perturbs the flow of the fluid~\citep{PhysRevLett.97.180405,amo2011polariton,PhysRevB.86.020509}. 

Polaritons possess a spin degree of freedom that arises from the two optically active exciton spin states which are coupled to the two circular polarizations of the light field. This spin structure gives rise to novel vortex states such as half-quantum vortices~\citep{PhysRevLett.99.106401,lagoudakis2009observation,PhysRevB.101.205301}, i.e., a vortex state in one spin component and a non-vortex state in the other. The energy splitting of perpendicularly polarized cavity photon modes~\citep{PhysRevB.59.5082}, known as photonic TE-TM splitting, leads to the spin-orbit interaction (SOI) of polaritons and many novel phenomena such as the optical spin Hall effect~\citep{PhysRevLett.95.136601,leyder2007observation} and oblique half-dark-solitons~\citep{PhysRevB.83.193305,hivet2012half}. It also leads to an interaction between vortices formed in different spin components and consequently influence their in-plane trajectories~\citep{dominici2015vortex,PhysRevB.91.085413}. The TE-TM splitting can induce an effective magnetic field in which half-solitons and half-vortices behave like ``magnetic charges'' with their propagation directions depending on the relative phase between the spin components~\cite{PhysRevB.85.073105,flayac2012separation}.

Condensates with opposite spins can also directly interact with each other, which induces an attractive nonlinearity~\citep{PhysRevB.82.075301}, even though its strength is typically much smaller than that of the intra-species repulsive interaction. It can be notably enhanced in a narrow spectral range close to the two-particle resonance associated with the formation of a bound biexciton state~\citep{PhysRevB.76.245324,takemura2014polaritonic}.

Here, we study the dynamics of vortices in ring lattices for polariton condensates imprinted by a spatially structured continuous wave optical pump beam. Due to the finite size of the pump spot, a steady condensate outward flow forms, building up a density gradient and pushing the existing vortices along the radial direction. Since the vortices are trapped in their radial position, the condensate flow results in an azimuthal orbit motion of the vortices due to the Magnus force~\citep{PhysRevB.55.485,fraser2009vortex}. In the spinor system, both the SOI and CI affect the orbital velocity of a single vortex. By slightly varying the curvature of the radial envelop of the periodic pump, the density gradient of the condensate along the radial direction can be adjusted, which also influences the size of the vortex core. When the SOI is significant and the condensates in both spin components are equally excited, i.e., under linearly polarized excitation, a new type of dark solution states, which are strongly deformed vortices along the azimuthal direction, can be found. In the presence of a weak CI, such kinds of dark solutions can form in only one spin component and behave like breathers, i.e. their sizes and angular velocities vary periodically during the propagation. When such dark solutions are formed in both spin components, they can even be frozen because of their interaction. As the curvature of the pump's spatial envelop increases, the dark solution can be further elongated along the azimuthal direction until occupying the whole ring. Interestingly, the phase defect carried by the dark solution then immediately triggers the snake instability that breaks the dark solutions in both spin components and leads to their decay. Besides affecting a single phase defect, the SOI also influences the collective circular motion of vortex constellations where the vortex-vortex and vortex-antivortex interactions become significant as discussed in detail below.

\begin{figure}[t]
  \centering
   \includegraphics[width=1.0\columnwidth]{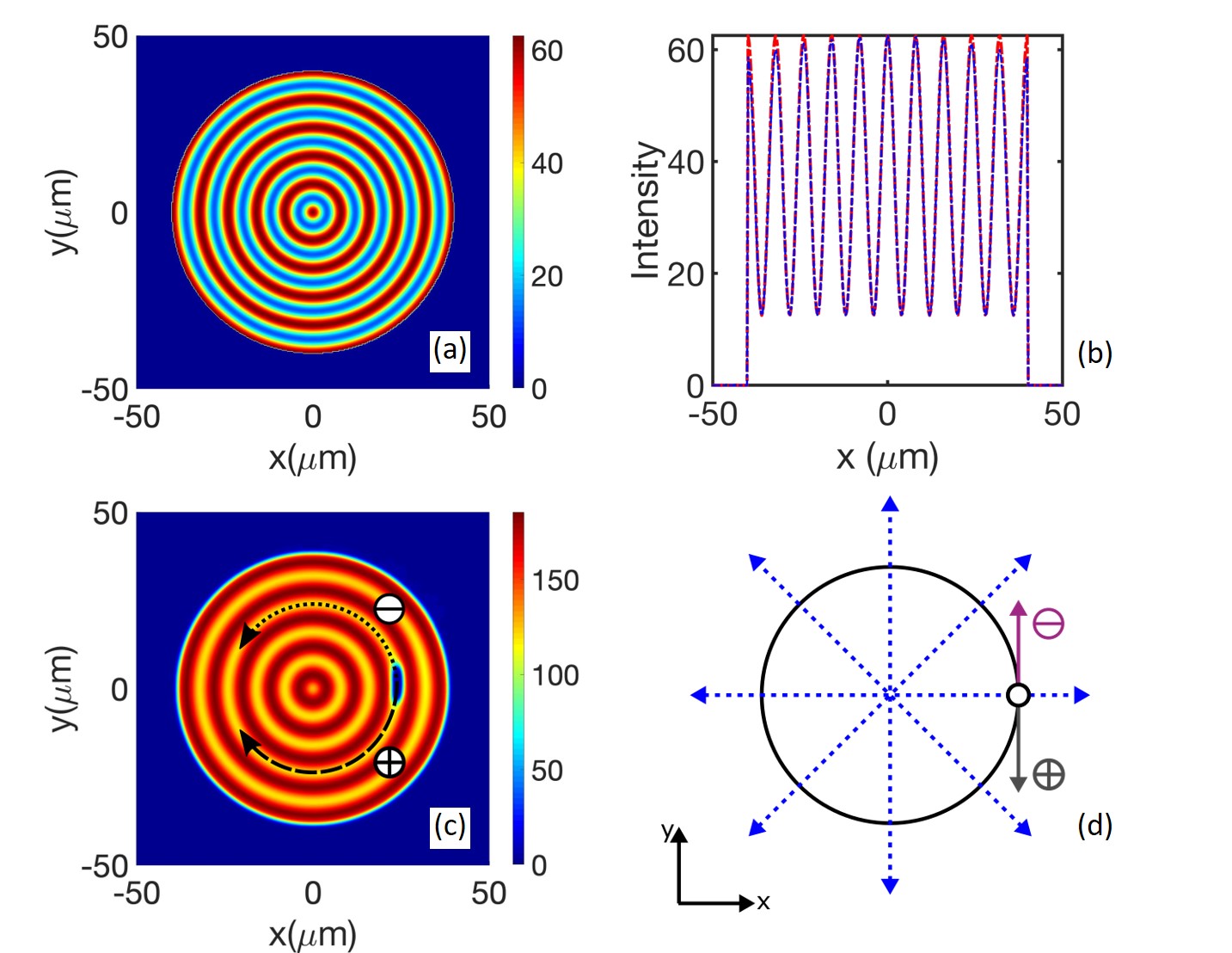}
  \caption{\textbf{Principle of vortex circulation.} (a) Radially modulated pump profile with intensity of $P_0=50~\mathrm{ps^{-1} \mu m^{-2}}$, orbit width of $d=8~\mathrm{\mu m}$, a maximum orbit radius of $r_\textup{c}=40~\mathrm{\mu m}$, and a flat envelop with $w\to\infty$. (b) The equivalent one-dimensional pump profile with $w\to\infty$ (red dashed line) and $w=100~\mathrm{\mu m}$ (blue dashed line). (c) Density of the condensate with a confined vortex inside, excited by the pump in (a). The arrows indicate the circulation directions of the vortex with a specific topological charge. (d) Principle of the Magnus force (purple and gray arrows), caused by the outward flow of the condensate (blue arrows), acting on oppositely charged vortices.}
  \label{fig:1}
\end{figure}

\section{Theoretical model}
To study the dynamics of binary polariton condensates under nonresonant excitation, we use the spinor driven-dissipative Gross-Pitaevskii model, including SOI caused by the TE-TM splitting and CI between different spin components~\citep{PhysRevLett.109.036404}, which reads
\begin{eqnarray}
\label{GP_psi}
    i\hbar\frac{\partial \psi_\pm}{\partial t} = \biggl(& - &\frac{\hbar^2}{2m_{\mathrm{eff}}}\nabla^2+g_\mathrm{c}|\psi_\pm|^2+g_\mathrm{x}|\psi_{\mp}|^2+g_\mathrm{r}n_\pm\nonumber\\& + & \frac{i\hbar}{2}[Rn_\pm-\gamma_\mathrm{c}] \biggr) \psi_\pm+J_\pm\psi_{\mp},
\end{eqnarray}
\begin{equation}
\label{GP_n}
    \frac{\partial{n_\pm}}{\partial{t}} = P_{\pm} - (\gamma_\mathrm{r}+R|\psi_\pm|^2)n_\pm.
\end{equation}
Here, $\psi_\pm=\psi_\pm(x,y)$ is the polariton field and $n_\pm=n_\pm(x,y)$ is the density of the reservoir. The subscripts $\pm$ denote the two spin components which correspond to right and left circular polarization. The two polariton fields with different spins are coupled to each other through $J_\pm= {\Delta}(\partial_x \mp i\partial_y)^2$, where ${\Delta}$ represents the strength of the TE-TM splitting. $m_{\mathrm{eff}}=10^{-4} m_\textup{e}$ ($m_\textup{e}$ is the free electron mass) represents the effective polariton mass. $g_\mathrm{c}=6~\mu$eV$\mu$m$^2$ describes the polariton-polariton interaction in the same spin, while $g_\mathrm{x}$ represents the CI of polaritons in opposite spins. $g_\mathrm{r}=2g_\mathrm{c}$ represents the interaction between the condensate and the reservoir in the same spin component. $R=0.01~\mathrm{ps^{-1}\mu m^2}$ denotes the condensation rate from the reservoir into the condensate and $\gamma_\mathrm{c}=0.2~\mathrm{ps}^{-1}$ characterizes the loss rate of the condensate. The loss rate of the reservoir is characterized by $\gamma_\mathrm{r}= 0.3~\mathrm{ps}^{-1}$. $P_{\pm}$ is the nonresonant pump whose profile can be tailored to achieve the desired spatial distribution of the condensate as well as the reservoir. In this work, to confine the phase defects and study their circular motion, we use a radially modulated continuous wave pump; see Fig. \ref{fig:1}(a), which is governed by 
\begin{equation}
\label{pump}
    P_{\pm}(\textbf{r}) = \left\{
\begin{array}{ll}
P_0\biggl( \mathrm{cos}^2\bigl(\frac{\pi\textbf{r}}{d}\bigr) \mathrm{exp}\left(-\frac{\textbf{r}^2}{2w^2}\right)+C_{0}\biggr), & |\textbf{r}| \leq r_\textup{c} \\
0. &  |\textbf{r}|>r_\textup{c} \\
\end{array}
\right. 
\end{equation}
Here, $P_0$ is the pump intensity, $d=$8 $\mathrm{\mu m}$ is the radial modulation constant, and $w$ represents the size of the pump spot and it also determines the curvature of the pump's envelop. For example, when $w\rightarrow\infty$, the envelop of the pump along the radial direction shows a flat top; see the red dashed profile in Fig. \ref{fig:1}(b). The curvature of the pump's envelop increases as the value of $w$ decreases; see the blue dashed profile in Fig. \ref{fig:1}(b). The compensation constant $C_{0}=0.25$ strengthens the intensity of the pump to modulate the contrast of the density distribution of the condensate to better confine the vortices. To create an outgoing flow of polaritons, we reshape the pump profiles by
cutting the outer rings at $|\textbf{r}| > r_\textup{c}$, where $r_\textup{c}$ represents the boundary of the pump. The finite number of the concentric rings in the pump is then given by $N_c=r_c/d-1$. Recently, it was found that Josephson vortices can be created in two concentric ring polariton condensates due to their complex coupling~\citep{PhysRevB.104.165305}. In our simulations, we also introduced a disorder potential with correlation length of about 1-2 $\mu$m and a mean depth of about 0.15 meV and found that all the results presented below remain unchanged, evidence that the dynamics presented are very robust.

\section{Single vortex circulation}

\begin{figure}[b]
  \centering
   \includegraphics[width=1.0\columnwidth]{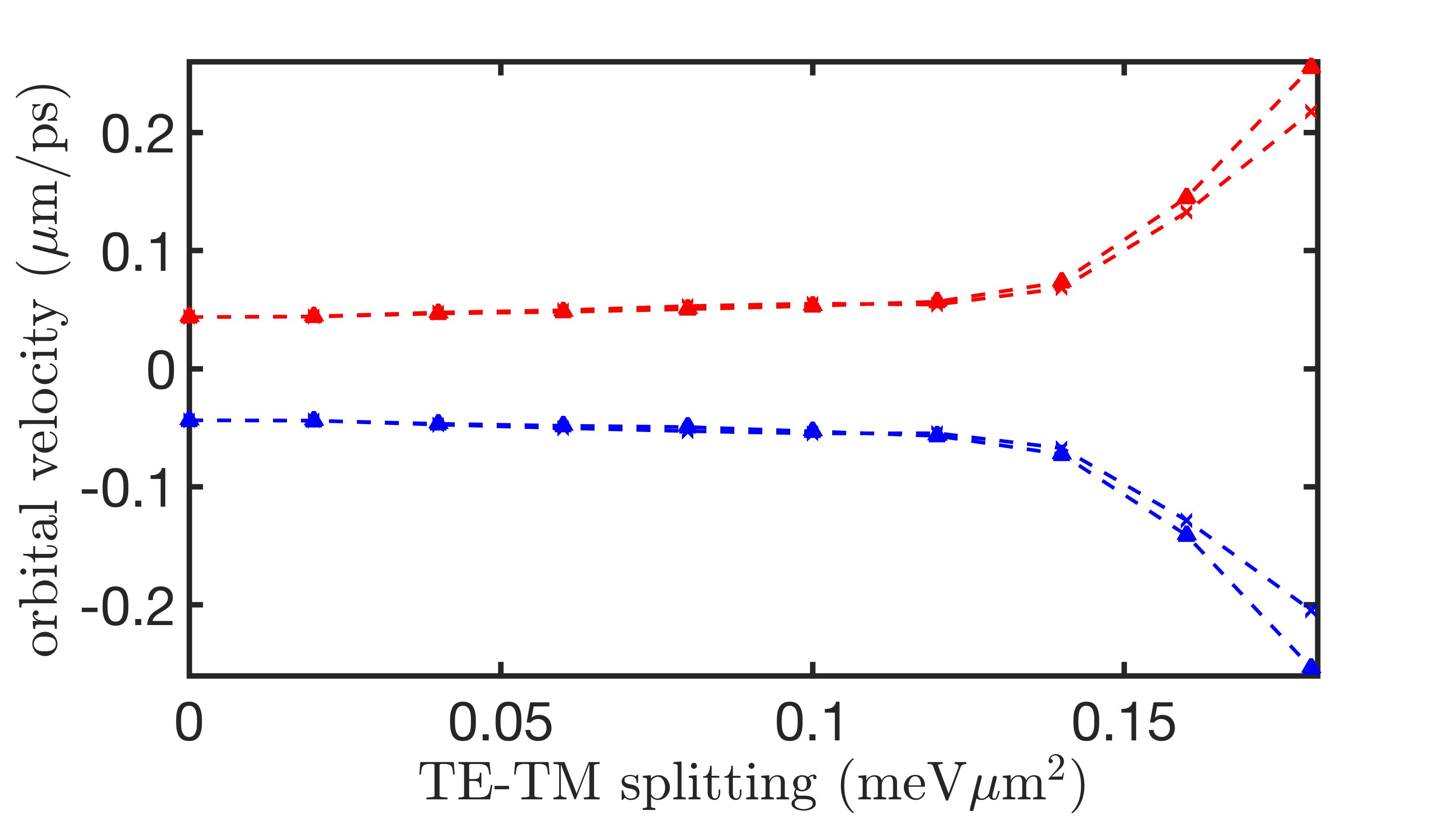}
  \caption{\textbf{Orbital velocities of vortices.} Orbital velocities of differently charged vortices ($m=+1$ for the blue lines and $m=-1$ for the red lines) on the third orbit from the center, in dependence of the TE-TM splitting for circularly polarized excitation. Additionally, crosses refer to a CI of $g_{\textup{x}}=0$ and triangles to $g_{\textup{x}}=-0.1~g_{\textup{c}}$.}
  \label{fig:2}
\end{figure}

Due to the condensate density differences within and outside the pumping region as well as the repulsive nonlinearity, polaritons spontaneously flow radially outward. This results in a density gradient inside the pumping area which also depends on the curvature of the pump’s spatial envelop, i.e. the larger the curvature, the faster the outflow.  A vortex can be initialized by placing a predefined phase defect at the desired target location in the corresponding vortex-free background solution of the system. Due to the effective potential caused by the radially modulated excitation, the vortex is trapped in the radial direction in a specific orbit. Importantly, the net radial outflow of the condensate causes a rotation of the vortex in its orbit due to the Magnus force. Note that the vortex does not circulate in its orbit if the Magnus force is absent. In other words, the vortex is pinned where it initially forms without the net outflow of the condensate. For a vortex with topological charge $m$ in a homogeneous condensate with density $n_c$ the explicit expression for the Magnus force reads~\citep{PhysRevB.55.485,fraser2009vortex}
    \begin{equation}
        \mathbf{F}_{\mathrm{M}}=mhn_c\mathbf{e}_z\times\mathbf{v}_{\mathrm{rel}}
    \end{equation}
where $\mathbf{v}_\mathrm{rel}$ is the relative velocity between the vortex and the surrounding condensate. In our case, an initially placed vortex inside a specific orbit starts to rotate either clockwise or counterclockwise depending on the sign of its topological charge. As the orbital velocity of the vortex increases the effective Magnus force becomes weaker since its component tangential to the orbit decreases. Finally, a stationary orbital velocity is reached indicating a balance of all forces acting on the vortex. The principle of the Magnus effect is illustrated in Fig. \ref{fig:1}(d) where we simply consider the scalar model of Eqs. \eqref{GP_psi} and \eqref{GP_n}, that is $\Delta=0$ and $g_\textup{x}=0$, under a pump intensity of $P_0=50~\mathrm{ps^{-1} \mu m^{-2}}$. For a counterclockwise rotating vortex, i.e. topological charge $m=+1$, the outward flow of the condensate [blue arrows in Fig. \ref{fig:1}(d)] results in a Magnus force, which is tangential to the orbit [grey arrow in Fig. \ref{fig:1}(d)], acting on the vortex. Conversely, if the topological charge of the vortex is $m=-1$, the direction of the Magnus force is flipped, as indicated by the purple arrow in Fig. \ref{fig:1}(d). In our system, the radially modulated condensate prevents the outgoing propagation of the vortices which are placed into the density valley of the background lattice, so that they can only circulate in the corresponding orbit and the circulation direction depends on the sign of the topological charge; see the arrows in Fig. \ref{fig:1}(c) and also the video in \citep{videos}.

\begin{figure}[b]
  \centering
   \includegraphics[width=1.0\columnwidth]{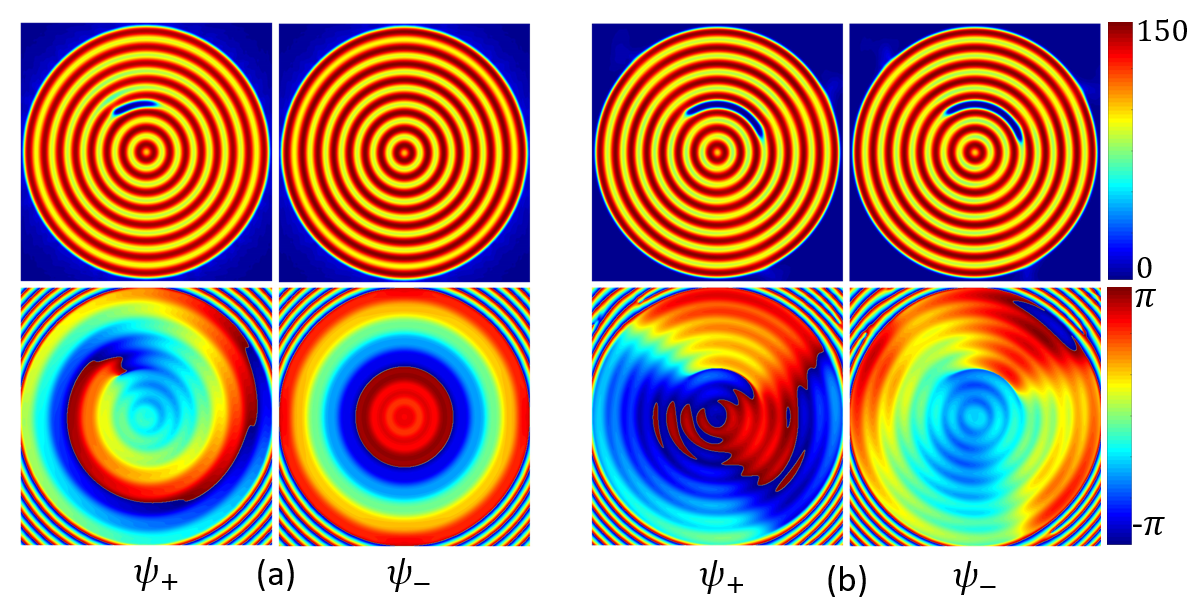}
  \caption{\textbf{Elongated dark solutions.} (a) Condensate density in $\mathrm{\mu m}^{-2}$ (upper row) and phase (lower row) distributions of the elongated half-vortex with an $m=-1$ charged vortex in the $\Psi_{+}$ component with $w=160~\mathrm{\mu m}$, $\Delta=0$, and $g_{\textup{x}}=0$. (b) Density in $\mathrm{\mu m}^{-2}$ (upper row) and phase (lower row) distributions of the frozen dark solution with $w=280~\mathrm{\mu m}$, $\Delta=0.1~\mathrm{meV\mu m^{2}}$, and $g_{\textup{x}}=-0.1$~$g_{\textup{c}}$. The spatial intervals shown range from $-80\,\mathrm{\mu m}$ to $+80\,\mathrm{\mu m}$ for all panels.}
  \label{fig:3}
\end{figure}

\begin{figure*}[t]
  \centering
   \includegraphics[width=2.0\columnwidth]{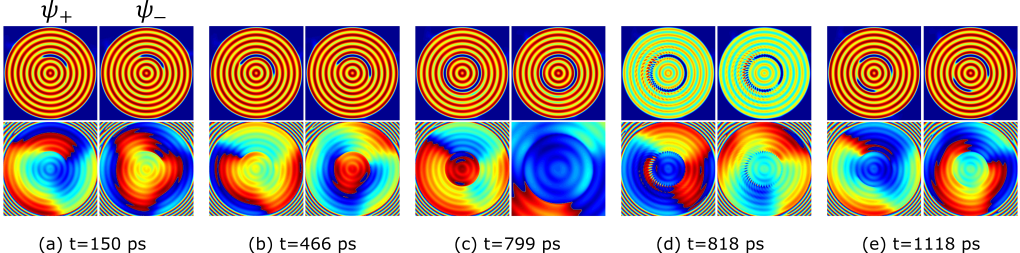}
  \caption{\textbf{Dynamic evolution of dark solution states.} Density in $\mathrm{\mu m}^{-2}$ (upper row) and phase (lower row) distributions of dynamic dark solutions at different times: (a) $t=150~\mathrm{ps}$, (b) $t=466~\mathrm{ps}$, (c) $t=799~\mathrm{ps}$, (d) $t=818~\mathrm{ps}$, and (e) $t=1118~\mathrm{ps}$. The window range for all the panels: -80$\sim$80 $\mu$m. A video showing the entire time evolution can be found in the Supplementary Material~\citep{videos}.}
  \label{fig:4}
\end{figure*}

The orbital velocity of a vortex is influenced by the TE-TM splitting and the CI as illustrated in Fig. \ref{fig:2} where the pump is circularly polarized with $P_{-}=0$. It shows the velocities of the vortices with different topological charges $m$ as a function of the TE-TM splitting $\Delta$. The orbital velocities are calculated as $v_o=2\pi R / T$ where $R$ is the orbit's radius and $T$ is the period. Both of the vortices with opposite topological charges can be accelerated by the TE-TM splitting; see the blue and red lines. Their velocities 
increase slowly and linearly with increasing SOI as long as $\Delta<0.12
~\mathrm{meV\mu m^{2}}$ and they are unchanged by introducing weak CI in this regime. The increase of their velocities is because the TE-TM splitting induces an effective magnetic field that accelerates differently charged vortices towards opposite directions~\citep{PhysRevB.85.073105,flayac2012separation}. However, when $\Delta>0.12~\mathrm{meV\mu m^{2}}$, their velocities become more sensitive to the TE-TM splitting as well as the weak CI. They increase drastically with the TE-TM splitting beause the larger TE-TM splitting leads to changes in the environment of the vortex, which influences the shape of the vortex core and consequently makes it feel a stronger Magnuce force. The CI further accelerates the circulating vortices. Larger velocities result in vortices not being trapped inside the orbits any longer; instead they completely escape from the pumping area.

\section{Breathing vortices and novel dark solutions}

Apart from the orbital velocities of the standard vortices being affected by the TE-TM splitting and the CI, their shapes and behavior almost remain unchanged under circularly polarized excitation. In this section we study the dynamics of vortices for linearly polarized excitation. In this case, due to the strong interaction of polaritons and vortices in different components, the vortices may escape from their orbits and consequently disappear at the edge of the excitation area. For a more efficient trapping of vortices in their own orbits and avoiding the influence from the edges, the radius of the excitation area is increased to $r_{c}=80~\mathrm{\mu m}$, resulting in $N_c=7$ concentric orbits as shown in Fig. \ref{fig:3}. Here, the confinement of the vortices is further enhanced by reducing the pump intensity to $P_0=45~\mathrm{ps^{-1} \mu m^{-2}}$ and increasing the modulation constant to $d=10~\mathrm{\mu m}$. Simultaneously, to strengthen the Magnus force, we slightly increase the curvature of the pump's envelop, that is the pump's envelop posses a more pronounced slope with $w=160~\mathrm{\mu m}$. Remarkably, the tilting of the pump elongates the cores of the vortices along only the azimuthal direction; see Fig. \ref{fig:3}(a). The size of this elongation is proportional to the curvature of the pump's envelop, but the curvature cannot be increased insignificantly, otherwise the outer rings becomes too weak to prevent the outgoing propagation of the vortices. It is worth noting that the elongation of the vortices is observed in the scalar case as well as under circularly polarized excitation.

When the TE-TM splitting is present and the CI is absent, the motion of the vortex in the $\Psi_{+}$ component can be perturbed by the non-vortex phase in the $\Psi_{-}$ component. As a result, the elongated vortex changes its orbital velocity and size periodically during the circulation, behaving like a breather; see the videos in \citep{videos}. Since the phase defect is only in one spin component, it can also be regarded as a breathing half-vortex. Including the CI enables the creation of the dark solution state in the $\Psi_{-}$ component as shown in Fig. \ref{fig:3}(b). We note that initially a vortex is imprinted in only the $\Psi_{+}$ component in Fig. \ref{fig:3}(b). Surprisingly, the dark solution in the $\Psi_{-}$ component reacts to that in the $\Psi_{+}$ component and further elongate the dark solutions in both components. If the TE-TM splitting is slightly increased, the dark solutions are further elongated to occupy a larger proportion of their respective orbits; see the video in \citep{videos}. These dark solution states are special in the sense that (i) they are frozen at the fixed position where they form, that is they do not move as time evolves, (ii) even though the dark state in the $\Psi_{+}$ component shows a 2$\pi$ phase winding, which is a typical property of a vortex, the phases between the two sides of the dark gap shows a clear $\pi$ phase difference, indicating a dark soliton, and (iii) circulation around the edge of the dark solution in the $\Psi_{-}$ component still gives a zero phase difference, i.e. there is no phase defect enclosed. However, there is still a $\pi$ phase shift between the two sides of the dark gap in the $\Psi_{-}$ component.

For the frozen solutions, the question arises whether such a dark solution can fully occupy its orbit. For this purpose, we slightly increase the curvature of the pump's envelop and the results are shown in Fig. \ref{fig:4}. It is clear that the two ends of each dark solution start to extend to occupy more of the orbit until it is completely filled; see Fig. \ref{fig:4}(a-c). After the dark solutions close and form a ring, it becomes a ring shaped dark soliton in the $\Psi_{+}$ component as shown in Fig. \ref{fig:4}(c). Dark solitons in 2D are unstable, such that they split into vortex-antivortex pairs in the homogeneous background~\citep{PhysRevB.89.235310} or show snake instability if they are confined along one direction~\citep{PhysRevLett.123.215301,Claude:20}. In our case, the dark soliton ring is also unstable, so that the snake instability is triggered immediately; see Fig. \ref{fig:4}(d). At the same time, the initillay imprinted phase defects in the $\Psi_{+}$ component drives the vortex-antivortex pairs to start to annihilate, leading to the decay of the dark solutions. Finally, only a part of the dark solution survives due to the initially imprinted phase defect; see Fig. \ref{fig:4}(e). Thereupon, the above process repeats robustly as time evolves; see the video in \citep{videos}.
During the evolution, the behavior of the dark solutions in both components are nearly synchronized. From Fig. \ref{fig:4}(c) one can see that the dark solution in the $\Psi_{-}$ component is quite different from a dark soliton, because of the clear $\pi$ phase difference at both sides of the density minimum vanishes.

\begin{figure}[t]
  \centering
   \includegraphics[width=1.0\columnwidth]{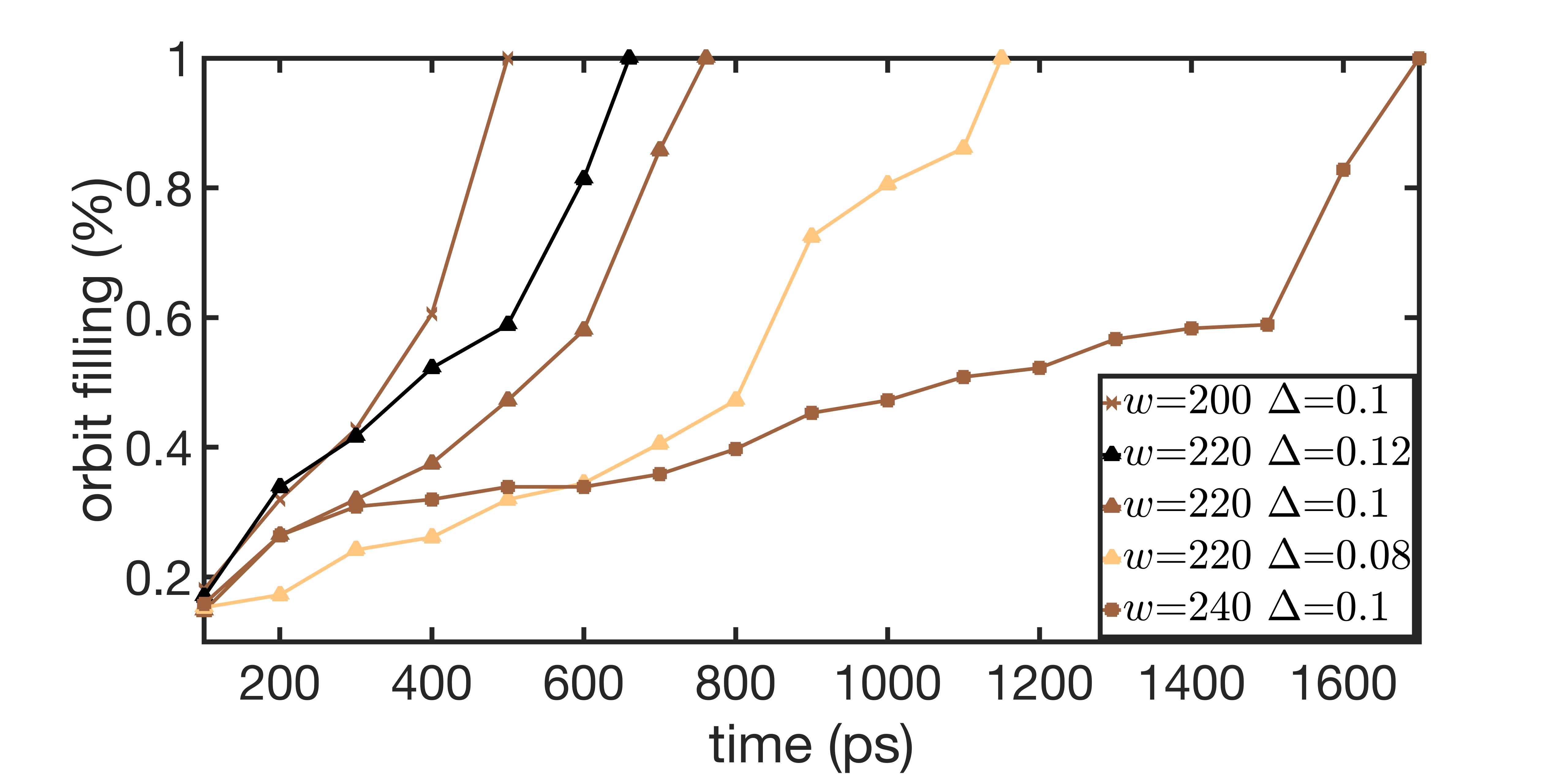}
  \caption{\textbf{Dark ring evolution.} Percentage of the third orbit filled by the dark solution as a function of time, in dependence of the TE-TM splitting $\delta$ and the curvature of the pump's envelope $w$.}
  \label{fig:5}
\end{figure}

In Fig. \ref{fig:5}, the orbit filling mechanism of the dark solution refering to Fig. \ref{fig:4}(a-c) is illustrated. One can see that the orbit filling velocity is associate with the TE-TM splitting and the curvature of the pump's envelope. For the same pump, a stronger TE-TM splitting accelarates the filling of the dark solution, while for the same TE-TM splitting, the filling mechanism becomes fast when the pump's envelope has a larger curvature. It can be seen that due to the appearance of the TE-TM splitting and the CI the orbit filling process is not linear over time.

\section{Collective motion of vortices}
\begin{figure}[t]
  \centering
   \includegraphics[width=1.0\columnwidth]{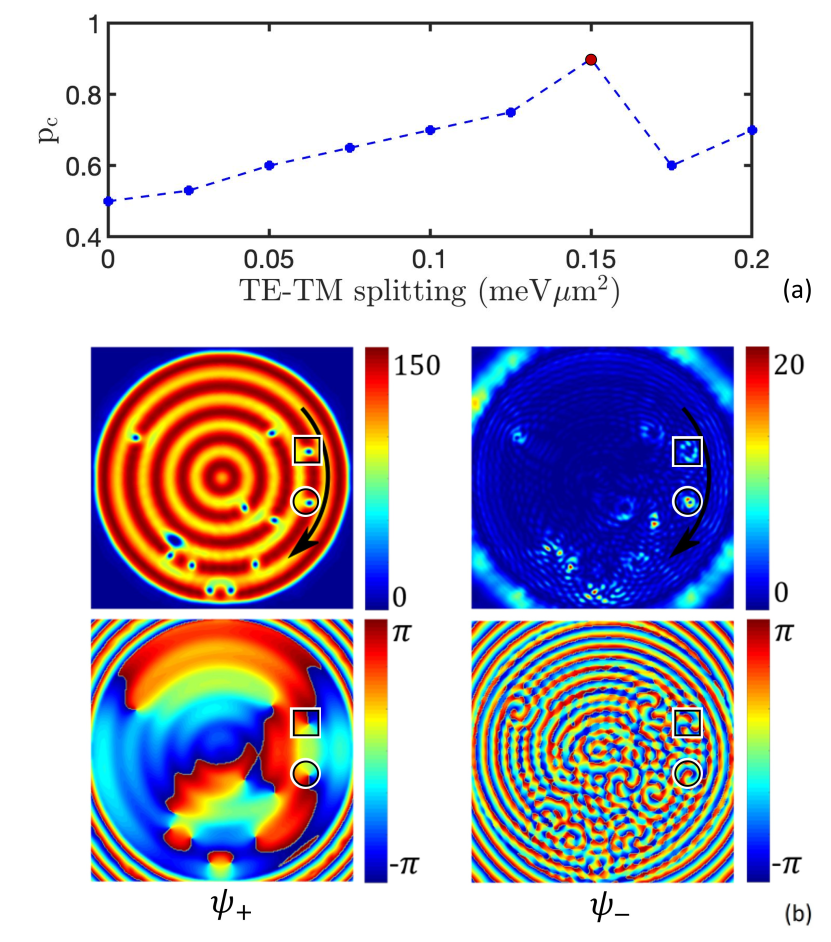}
  \caption{\textbf{Rotation of vortex constellations.} (a) Probability of the clockwise rotation $p_c$ of the constellation in the $\Psi_{+}$ component depending on the TE-TM splitting. (b) Density in $\mathrm{\mu m}^{-2}$ (upper row) and phase (lower row) distributions of a vortex constellation at $\Delta=0.15~\mathrm{meV\mu m^{2}}$, corresponding to the red point in (a) under a pump intensity of $P_0=50~\mathrm{ps^{-1} \mu m^{-2}}$. The arrows indicate the circulating direction. The window range for all the panels: -40$\sim$40 $\mu$m. A video showing the time evolution can be found in the Supplementary Material~\citep{videos}.}
  \label{fig:6}
\end{figure}

In this section, we study the collective motion of multiple phase defects trapped in different orbits. To this end, we use white noise as initial condition and a circularly polarized pump. In this case, multiple vortex-antivortex pairs can form from the initial noise, and the interaction between different pairs results in the formation of a fixed vortex constellation. After the constellation stabilizes, it starts to circulate as a whole, either clockwise or counterclockwise. Without TE-TM splitting and CI, the probabilities for the clockwise rotation $p_{c}$ and the counterclockwise rotation $p_{cc}=1-p_{c}$ are equal, i.e. $p_{cc}=p_{c}=0.5$, as illustrated in Fig. \ref{fig:6} (a). These probabilities are derived by determining the proportions of clockwise and counterclockwise rotating constellations and each probability is calculated by averaging over $40$ simulations 
by using a different noise realization as initial condition for each simulation. 
Note that the spontaneous formation of the vortices from noise results in that they can build up either in the density valleys or at the density peaks~\citep{PhysRevLett.118.157401}. Since the vortices in the density valleys have larger sizes, so that their interaction leads to recombination for most of them before the group motion establishes, more vortices survive in the density peaks with smaller sizes.

Turning on the TE-TM splitting shifts the probability of the rotation directions of the constellation. A right (left) circularly polarized pump increases the probability of the collective clockwise (counterclockwise) movement in the presence of the TE-TM splitting. Figure \ref{fig:6} (a) shows the possibility of the clockwise rotation $p_{c}$ in the $\psi_+$ component in dependence of the TE-TM splitting. For a value of $\Delta=0.15~\mathrm{meV\mu m^{2}}$ the probability gets quite close to $p_c=1$, whereas the probability decreases as the TE-TM splitting increases further. The reason is that at larger TE-TM splitting, the density of the condensate in the $\psi_-$ component is enhanced, so that it strongly affects the vortex and the surrounding environment in the $\psi_+$ component. Another reason is that the stronger TE-TM splitting induces a prominent effective magnetic field which leads vortices that carry different topological charges to propagate in opposite directions~\citep{PhysRevB.85.073105,flayac2012separation}. Consequently, the nearly unidirectional circulation of the vortex constellation is broken.

Besides affecting the collective motion of the vortices, the SOI also induces the formation of bright vortices or localized vortices in the $\psi_-$ component where the pump is inactive as shown in Fig. \ref{fig:5}(b). One can see that each smaller vortex in the $\psi_+$ component corresponds to a bright vortex in the $\psi_-$ component, and their topological charges satisfy the relation $|m_+-m_-|=2$, agreeing with the relation of two coupled non-localized vortices~\citep{PhysRevB.81.045318}. For example, in the $\psi_+$ component the smaller vortex with topological charge $m_{+}=1$ ($m_{+}=-1$) has a bright counterpart at the same position in the $\psi_-$ component with topological charge $m_{-}=-1$ ($m_{+}=-3$), as marked by the circles (rectangles) in Fig. \ref{fig:6}(b). Note that the circulating directions of the vortex constellations in different spin components are synchronized due to the phase coupling, originated from the SOI, of the paired vortices in different spin components. The association of these two kinds of vortices in different spin components provides a method to control the bright vortices by manipulating the dark vortices~\citep{PhysRevLett.118.157401} in the other spin component.

\section{Conclusion}
In summary, we have studied the dynamics of phase defects trapped in concentric rings in spinor polariton condensates. We find that for circularly polarized excitation, the single vortex circulatory motion, driven by the Magnus force, can be influenced by the CI and the SOI. The collective motion of vortex constellations can also be affected by the SOI. Both CI and SOI play a crucial role for linearly polarized excitation, especially for the creation of the novel frozen dark solution states. We find that the size of the frozen dark solution is related to the curvature of the pump's spatial envelop. In some cases, the frozen solutions can even occupy the whole ring, which is an unstable scenario, and then a snake instability is triggered. Our results demonstrate control of phase defects through SOI and CI, which may be of interest to the conservative atomic condensates, nonlinear optics, and other binary physical systems.

\begin{acknowledgments}
This work was supported by the Deutsche Forschungsgemeinschaft (DFG) through the collaborative research center TRR142 (grant No. 231447078, project A04) and by the Paderborn Center for Parallel Computing, PC$^2$. X.M. further acknowledges individual grants from the DFG (No. 467358803) and NSFC (No. 11804064).
\end{acknowledgments}

\end{document}